# *Anomalous Hardening in Magnesium Driven by a Size-Dependent Transition in Deformation Modes*

Gi-Dong Sim[1], Gyuseok Kim[2], Steven Lavenstein[1], Mohamed H. Hamza[1], Haidong Fan[1,3], Jaafar A. El-Awady[1, *]

[1]Department of Mechanical Engineering, Johns Hopkins University, Baltimore, MD 21218, USA

[2]Singh Center for Nanotechnology, University of Pennsylvania, Philadelphia, PA, 19104, USA

[3]Department of Mechanics, Sichuan University, Chengdu, Sichuan 610065, China

[*] Email: jelawady@jhu.edu

**Abstract**

Magnesium (Mg) and its alloys hold great potential as an energy-saving structural material for automative, aerospace applications. However, the use of Mg alloys has been limited due to poor ductility and formability. Poor mechanical properties of Mg alloys origin from the insufficient number of slip systems, and deformation twinning plays an important role to accommodate plastic deformation. Here, we report a comprehensive experimental and modeling study to understand crystal size effect on the transformation in deformation modes in twin oriented Mg single crystals. The experiments reveal two regimes of size effects: (1) single twin propagation, where a typical "smaller the stronger" behavior was dominant in pillars ≤ 18 μm in diameter, and (2) twin-twin interaction, which results in anomalous strain hardening in pillars > 18 μm. Molecular dynamics simulations further indicate a transition from twinning to dislocation mediated plasticity for crystal sizes below a few hundred nanometers. Our results provide new understanding of the fundamental deformation modes of twin oriented Mg from nano-scale to bulk, and insights to design Mg alloys with superior mechanical properties through dimensional refinement. This subsequently can materialize into more utilization of Mg alloys as a structural material in technologically relevant applications.

Magnesium (Mg) and its alloys have been garnering significant interest as structural materials for many technologically relevant applications in the automative, aerospace, electronics, and biomedical industries, due to the their lightweight, high strength, and superior damping capacity[1-4]. However, the use of Mg alloys as a structural material has been limited to date due to their poor ductility and formability at room temperature, as well as their poor creep resistance at elevated temperatures[1, 5]. The poor ductility of Mg originates from the hexagonal closed packed (HCP) lattice structure, which exhibits low crystal symmetry. As a result, Mg does not have sufficient active slip systems at room temperature, which dictates that deformation twinning plays an important role in accommodating plastic deformation[6]. Thus, quantifying the competitive nature between slip and twinning is necessary to understand the deformation behavior of Mg and subsequently improve its properties through alloying.

Over the past decade, the effects of intrinsic and extrinsic length scales on the deformation mechanisms of micron sized single crystals[7-12] and polycrystalline thin films[13-22] have drawn considerable attention. While deformation twinning in HCP crystals is common and is a highly competitive deformation process along with dislocation slip over a broad range of length scales, the effect of the crystal size on these competitive deformation processes is still relatively unexplored. Only a limited number of studies on polycrystalline Mg[23] and Mg alloys[24-26] reported the existence of a critical grain size on the order of a few microns, below which deformation twinning is suppressed and dislocation mediated plasticity predominantly governs plastic deformation. In addition, recent discrete dislocation dynamics (DDD) simulations of the competition between dislocation slip and $\{10\bar{1}2\}$ tension twinning in Mg[27] suggested that twinning deformation exhibits stronger grain size effects as compared to dislocation slip in polycrystalline Mg, which is the main reason for the change in deformation mode at a critical grain size. It should be noted that polycrystalline Mg and Mg alloys are more technologically relevant over single crystal Mg. However, experimental results on polycrystalline materials can be ambiguous to analyze due to the complicated crystal orientation, variation in grain size, and influence from grain boundaries. On the other hand, single crystal Mg and Mg alloys experiments can provide a detailed quantification of the competitive nature of twinning and dislocation slip that would provide insights to better fundamental understanding of the deformation mechanisms in polycrystalline Mg and Mg alloys.

Furthermore, while a small number of size effect studies on single crystal Mg and Mg alloy microcrystals have been reported recently, most were limited to c-axis compression that is not a favorable orientation for twinning, or to microcrystal having sizes in the range of a few microns[28-31]. In these later studies, the twinning deformation mode was governed by a single twin variant that leads to strain softening or massive strain burst[29, 32-36]. These observations are characteristically different than that for bulk Mg and Mg alloys, were multiple twins with different variants are activated and interact with each other, leading to considerable strain hardening[37-39]. Thus, understanding the transition in twin modes may provide powerful insights to tailor the mechanical behavior of Mg alloys through dimensional refinement to facilitate the future use of Mg alloys as a lightweight structural material.

In the present study, the transition in deformation modes as the crystal dimension changes in Mg single crystals is fully characterized. Micro-compression experiments are performed on single crystal Mg microcrystals ranging in size from 2.0 to 22.6 μm and compared with bulk scale results of the same orientation. Two regimes with different twin modes are observed - 1) pillars with diameter below 18 μm, where single twin propagation dominates and 2) pillars above 18 μm, where twins with different variants start to interact. Molecular dynamics (MD) simulations are also performed and indicate a transition from twinning to dislocation mediated plasticity for crystal sizes below a few hundred nanometers. The size-affected response on the flow strength for the different deformation modes are discussed, and a deformation mechanism map for twin oriented Mg single crystal is predicted based on current observations.

**Deformation modes during compression along the a-axis**

The critical resolved shear stress (CRSS) measured experimentally[40-46], the Peierls stress predicted from MD simulations[47-52], and the Schmid factor for all possible deformation modes in Mg for the compression loading along the $[11\bar{2}0]$ in the matrix, and $[2\bar{1}\bar{1}\bar{3}]$ in the twinned region are summarized in Table I. For the $[11\bar{2}0]$ imposed compression loading, the Schmid factor is highest for prismatic slip in the matrix (i.e. parent crystal), followed by $\{10\bar{1}2\}$ tensile twinning. However, the CRSS for prismatic slip at room temperature is four times higher than that for tensile twinning. Thus, deformation is most likely dominated by tension twinning. Within the twined region, basal slip is expected to dominate due to its low CRSS and high Schmid factor.

Representative engineering stress-strain curves for microcrystals having $D \leq 18$ μm are shown in Fig. 1(a), while those for $D > 18$ μm are shown in Fig. 1(b). In addition, the engineering stress-strain response of a bulk Mg single crystal compressed along the $[11\bar{2}0]$ direction[53] is also shown in Fig. 1(b) for comparison. It is observed that microcrystals having D ≤ 11 μm show massive strain bursts that commence at about 1% strain, and while strain bursts are still observed for microcrystals having $D > 11$ μm, their magnitudes are significantly smaller. In addition, no significant strain hardening is observed for crystals having $D \leq 18$ μm. It should be noted that these observations are in agreement with previous single crystal Mg pillars experiments in the range of ~ 0.5 > D > 11 μm [33, 34]. On the other hand, microcrystals having D > 18 μm show significant strain hardening, which is characteristically different than smaller crystals or the bulk response. This strain hardening is observed to gradually decrease toward the bulk response with increasing crystal size.

**Microcrystals smaller than 18 μm (single twin dominated response)**

The SEM micrographs of a microcrystal having $D = 10$ μm pre- and post-deformation (after 15% strain) are shown in Fig. 2(a), and 2(b), respectively. These morphologies are representative for all microcrystals having $D \leq 18$ μm. To further elucidate the deformation mechanisms in these microcrystals, ion beam channeling contrast and EBSD images of a thin lamella extracted longitudinally from the center of a $D = 10$ μm microcrystal after ~5% strain using the FIB lift-out technique are shown in Fig. 2(c) and 2(d), respectively. These micrographs indicate that a single twin spans most of the top half of the column. The loading axis in the twinned region is parallel to the $[2\bar{1}\bar{1}3]$ direction, which leads to a misorientation angle between the twin and matrix of approximately 86° between the twin and matrix. Twins are observed to nucleate from the top of the pillar[33, 34], then propagate laterally to form a thin twin lamella, as shown schematically in Supplementary Fig. S1(a). Once a complete twin lamella forms, it thickens following a layer-by-layer migration throughout the column of the pillar.

It should be noted that during compression along the $[11\bar{2}0]$ direction the Schmid factor is maximum for two twin-variants, namely, $(\bar{1}012)[\bar{1}011]$ and $(0\bar{1}12)[0\bar{1}11]$ twins. Thus, the probability of the nucleation of either twin variant is equal, and twin-twin interaction is generally expected. However, as shown in Fig. 2(d) only a single twin-variant is observed in all tested Mg microcrystals having $D \leq 18$ μm. This single twin dominated deformation for the $D \leq$ 18 μm can be attributed to a number of factors including: 1) the surface roughness of the microcrystals; 2) the friction between the microcrystal and the flat punch surfaces; and/or 3) the microcrystal taper angle that leads to higher stresses at the top surface. Since the twin nucleation stress is higher than the twin propagation stress[54] and the twin tip velocity is high, when a twin nucleates at the loading surface it propagates down the column before another twin can nucleate. In this scenario, other possible sites where additional twins can nucleate and propagate are not in the column of the pillar but within the contact area.

Within the twined region, basal slip is expected to dominate due to its low CRSS and high Schmid factor (see Table 1). This is confirmed in the current experiments by the observed basal slip bands in the twinned region, as shown in Fig. 2(b) and 2(c). The single slip character of basal slip leads to no measurable strain hardening. As shown schematically in Fig. 2(e), the loading axis with respect to the twin crystallography is $[2\bar{1}\bar{1}3]$ and the angle between the loading axis and the basal planes in the twin is 31.6°. Given that the aspect ratio of the pillars is 2:1, the twin thickness must be at least 60% of the pillar length before basal slip bands can extend from one side of the pillar to the other (i.e. twin volume fraction of 0.6). The longitudinal plastic strain required to reach this condition can be predicted from[33, 34]

$$\varepsilon = VM\gamma \qquad (1)$$

where $V$ is the twin volume fraction, $M = 0.37$ is the Schmid factor for $[11\bar{2}0]$ loading, and the shear strain due to twinning is $\gamma = 0.129$[6]. Thus, a longitudinal plastic strain of at least 2.9% is required before basal slip bands are observed on the microcrystal surface.

The smaller strain burst observed for crystals having 11 μm $< D \leq$ 18 μm as compared to those observed for crystals having $D \leq 11$ μm could be explained as follows. In the smaller microcrystals the flow stress levels are high as compared to those in the larger microcrystals.

Since the twin-boundary velocity is proportional to the stress-level[55], the twin-boundary velocity, and subsequently the plastic strain rate, in the smaller crystals will be higher. In addition, the nanoindenter's feedback-control loop has a fixed speed, which dictates how fast the system can respond when a strain burst commences. In case the strain burst due to twin-boundary propagation reaches the critical plastic strain for the onset of basal slip band formation in the twin before the nanoindenter responds to the burst, additional strain burst would occur due to basal slip. This is clear for crystals having D ≤ 11 μm, which show strain bursts larger than the 2.9% strain required for the twin to reach the critical volume fraction for basal slip bands to be observed. For crystals having 11 μm < $D$ ≤ 18 μm, the strain bursts are smaller than 2.9%, which indicate that the nanoindenter is able to retract shortly after the onset of strain bursts.

**Microcrystals larger than 18 μm (Twin-twin interaction induced hardening)**

A representative SEM image of a microcrystal having $D$ = 20 μm after 8% strain, is shown in Fig. 3(a). Ion beam channeling contrast and EBSD images of a thin lamella extracted from the center of the upper half of the same microcrystal, are shown in Figs. 3(b) and 3(c), respectively. Unlike microcrystals having $D$ < 18 μm, it is observed that deformation is dominated by the nucleation and interactions of multiple twin-variants. As shown schematically in Supplementary Fig. S1(b), two twins having different variants nucleate then propagate before either twin is able to form a complete twin lamella in the microcrystal. This inhibits the formation of clear basal slip bands in the twined regions. Three-dimensional (3D) DDD simulations have shown that the twin boundaries act as strong obstacles to basal slip, and contribute significantly to the overall hardening response[56]. Thus, when multiple twin-variants nucleate and interact, subsequent basal slip in either twinned region will be constrained by the twin boundaries, which leads to significant strain hardening as shown in Fig. 1(b).

It should be noted that the hardening rate is the highest in the smallest microcrystals that show twin-twin interactions, and decreases as the microcrystal size increases approaching the hardening rate of a bulk Mg crystal. Since hardening is due to twin-twin and dislocation-twin interactions, the effect is mostly localized to the area near the twin boundary. Therefore, the effect on the macroscopic stress-strain curve will decrease as the pillar-size, and subsequently the twin-size, increase until the bulk response is recovered in large enough microcrystals. These observations are in agreement with DDD simulations that show a strong twin size-effect on the hardening of a-axis oriented microcrystals[57].

**Size effects and the deformation mechanisms map**

The critical resolved shear stress ($\tau_{CRSS}$) calculated from the flow stress ($\sigma_f$) at 4% strain ($\tau_{CRSS} = \sigma_f \cdot M$) versus the microcrystal diameter for all the current experiments are shown in Fig.4(a). Here $M = 0.37$ is the Schmid factor for [11$\bar{2}$0] loading when deformation is predominantly tension twinning, while $M = 0.43$ for [11$\bar{2}$0] loading when the deformation is predominantly prismatic slip only. Two distinct regions with different size-affected responses can be clearly observed for microcrystals above and below $D = 18$ μm. In the first regime (D ≤ 18 μm), the size effects are observed to follow a power-law relationship of the form $\tau_{CRSS} \propto D^{-0.77}$. This power-law exponent ($n = 0.77$) is smaller than that for tensile twinning during [10$\bar{1}$0] tensile-loading ($n = 0.93$)[34], and that predicted from the stimulated slip model for twin growth ($n = 1$)[32]. The lower exponent in the current experiments can be rationalized based on the two types of active dislocations in this loading axis, namely, prismatic dislocations and twinning dislocations[27]. Even though plastic strain mediated by twin growth is significantly larger than prismatic slip, prismatic dislocations are expected to affect twin growth by dislocation-twin boundary interactions[27, 56]. It should be noted that the power law exponent for tensile twinning as predicted from 3D-DDD simulations ($n = 0.75$)[27], which considers dislocation-twin boundary interaction, surprisingly matches well the current experimental measurements ($n = 0.77$). Nevertheless, while the DDD simulations focused on interactions between the twin boundary and prismatic dislocations in the matrix, interactions between the twin boundary and basal dislocations within the twinned region might be another important factor to consider.

In the second regime ($D > 18$ μm), a sudden increase in strength is observed resulting from twin-twin and dislocation-twin interaction. As discussed earlier, the hardening rate gradually decreases toward the bulk behavior as the microcrystal size further increases. As such, the strength decreases with increasing microcrystal size following another power-law relationship. However, the range of microcrystal sizes tested is small due to the significant time needed to FIB mill larger pillars. Thus, the magnitude of the power law exponent in this region was not quantified here.

On the other hand, the stress-strain response and the deformation mode as predicted from MD simulations for $30 \times 30 \times 60$ nm$^3$ and $100 \times 100 \times 100$ nm$^3$ simulation cells are shown in Supplementary Figs. S2 and S3, respectively, and the evolution of deformation microstructure in the $30 \times 30 \times 60$ nm$^3$ simulation cell is shown in Supplementary Movie S1. In both cases prismatic <a> dislocations are observed to nucleate on the two prismatic planes having the highest Schmid factor. These prismatic <a> dislocations carry the plastic strain throughout the deformation, and no twinning deformation was observed in either simulation cell up to the end of the simulations (> 15%).

The critical resolved shear stress at yield versus the simulation cell edge length for the two simulation cells are also included in Fig. 4(a). It is clear that the slope predicted by these results for dislocation slip is considerably smaller than the slope predicted from the current experiments for deformation with twinning. Through extrapolation, both curves intersect at a crystal size of ~120 nm, which suggests that below this crystal size deformation would be dominated by dislocation slip, while that above it deformation will be dominated by twinning. To confirm this, a cross-sectional view through a 140 × 140 × 70 nm$^3$ simulation cell after 8.15% strain is shown in Fig. 4(b), while the 3D microstructure is shown in Supplementary Fig. S3(c). For this crystal size, prismatic <a> dislocations are observed to nucleate first on the two prismatic planes having the highest Schmid factor. However, at 5.6% strain twin nucleation initiates at a dislocation junction that forms near the free surface, which grows slowly with increasing deformation. From the orientations of the matrix and twin shown in the insert of Fig. 4(b), the c-axis of the twinned region is rotated by ~90° from the matrix, indicating it is a $\{10\bar{1}2\}$ tension twin, in agreement with those observed in the current experiments. Furthermore, two types of interfaces between the matrix and the twin can be identified, including straight basal/prismatic (B/P) transformation interface, and curve TBs, showing good agreement with the experimental TEM and MD observations[51, 58]. These MD simulations confirm the transition from predominant dislocation mediated plasticity to twinning dominated plasticity at a critical crystal size, as shown in Fig. 4(a). Due to the high strain rates employed in MD simulations as compared to those in the current experiments, it is expected that the yield strength would decrease, and subsequently the critical crystal size will increase, with decreasing strain rate[59], as shown schematically in Fig. 4(a).

The question that arises here is what are the physical origins that control the transition between the different deformation modes discussed above. To answer this it is important to note that twin nucleation is governed by the local stress state near pre-existing defects (e.g. grain boundaries, surfaces, and dislocations), while twin propagation is governed by long range stresses. Furthermore, the origin of the variability in twinning modes is mostly attributed to twin nucleation probability[60, 61]. While various models for homogeneous[62] and heterogeneous[63-66] twin nucleation have been proposed in literature, homogeneous twin nucleation is less likely to occur due to the extraordinarily high stress required. Alternatively, heterogeneous twin nucleation at a pre-existing defects is more likely to play a dominant role in governing the twin nucleation process. In heterogeneous nucleation models, dissociations of perfect <a>, <a+c>, or <c> dislocations into glissile twinning dislocations has been suggested as the main nucleation mechanism for twin nucleation in Mg[64, 65]. These models agree with recent atomistic simulations, which show that it is energetically more favorable to nucleate a twin in an HCP crystal by the simultaneous nucleation of a partial dislocation and multiple twinning dislocations[67, 68] from pre-existing defects.

In the current Mg single crystal experiments, sessile screw dislocations may act as local stress concentrators that promote twin nucleation. When the applied stress is high enough to activate both prismatic edge and screw dislocations (i.e. $\tau_{applied} > \tau_{PSscrew} > \tau_{PSedge}$), prismatic dislocations will glide continuously and plastic deformation will be governed by prismatic slip (see schematic in Supplementary Fig. S4(a)). Since the Peierls stress for prismatic edge dislocations are lower than screw dislocations (Table I), if the applied stress is higher than the CRSS for edge dislocations but lower than the CRSS for screw dislocations, edge dislocations will be mainly active, which will lead to the formation of sessile screw dislocations in the crystal (see schematic in Supplementary Fig. S4(b)).

For a typical initial dislocation density on the order of $10^9$-$10^{12}$ m$^{-2}$, in submicron crystals (D < 0.5 μm) the applied stress is high and the crystal size is small, thus, both screw and edge dislocations will be activated and the probability of twin nucleation as compared to dislocation mediated plasticity greatly decreases. For crystals in the range of $0.5 \leq D \leq 18$ μm, the number of available sources to create sessile screw dislocations is typically small; therefore, the probability to nucleate multiple twin-variants is small. On the other hand, in larger crystals ($D \geq 18$ μm), more dislocation sources are available, which results in increasing the number of probable sites for twin nucleation (see schematic in Supplementary Fig. S4(c)).

In summary, micro-compression room temperature experiments were performed on single crystal Mg microcrystals ranging in size from 2 μm to 22.6 μm. Two regimes with different twin modes are reported. In the first regime (diameters smaller than 18 μm) single twin propagation dominates, while in the second regime (diameters larger than 18 μm) nucleation and interactions between different tensile twin-variants are observed. The stress required for twin propagation was found to increase with decreasing sample size, showing a typical "smaller the stronger" behavior. An anomalous increase in strain hardening is first reported for microcrystals having diameters larger than 18 μm, which is induced by twin-twin and dislocation-twin interactions. The hardening rate gradually decreases toward the bulk response as the microcrystal size further increases. Furthermore, molecular dynamics simulations indicate a transition from twinning mediated plasticity to dislocation mediated plasticity for crystal sizes below a few hundred nanometers in size. A deformation mechanism map for twin oriented Mg single crystals, ranging from the nano-scale to bulk scale is proposed based on the current simulations and experiments. The current predicted size-affected deformation mechanism of twin oriented Mg single crystals can lead to better understanding the competition between dislocations plasticity and twinning plasticity. In addition the current results can further be used to validate physics-based simulations methods (e.g. atomistic simulations, discrete dislocation dynamics simulations, and crystal plasticity) targeted at modeling and predicting the different mechanical properties of Mg and Mg alloys. While the deformation mechanisms map proposed here is based on tensile-twin oriented Mg single crystals, the general trend of length scale effect on the dominant deformation mechanism should be applicable to other HCP metals.

**Methods**

**Sample fabrication**

An a-axis [11$\bar{2}$0] oriented single crystal Mg rod, purchased from Metals Crystals and Oxides Ltd. UK, was cut using wire electrical discharge machining (EDM) to extract a 12 mm diameter disk having 3 mm thickness. The disk was then chemically polished in a solution of 10% nitric acid in deionized (DI) water, followed by electropolishing using a solution of 5% nitric acid in methanol electrolyte, to remove the oxide and EDM induced damage surface layers. The initial dislocation density was measured using the etch-pitting technique by exposing the surface of the sample to a solution consisting of 10 g ammonium chloride dissolved in 50 cm$^3$ of DI water[69, 70]. The estimated initial dislocation density was on the order of $10^9$ m$^{-2}$.

Magnesium [11$\bar{2}$0] oriented cylindrical pillar-like microcrystals, ranging in diameter from $D$ = 2.0 to 22.6 μm, were then milled into the disk using the FIB annular milling technique. The microcrystals' aspect ratio were approximately 2:1 (height to mid-plane diameter) to avoid buckling for higher aspect ratios, or non-uniform stresses along the length for lower aspect ratios[71]. The taper angle resulting from the annular milling process was approximately 2º, which is not expected to affect the deformation mechanisms of interest in the current study.

**Mechanical testing**

All microcrystals were tested at room temperature in uniaxial-compression[7] using an *in situ* scanning electron microscope (SEM) nanoindentation setup (InSEM, Nanomechanics Inc.), equipped with a 30 × 30 μm diamond flat tip punch. The *in situ* SEM setup reduces the possibility of surface oxidation during the mechanical test. While the instrument is inherently load controlled, the displacement rate was controlled through the feedback from the load signal in order to achieve a nominal strain rate of $10^{-3}$ $s^{-1}$. If a displacement jump larger than 10 nm was recorded during the deformation (e.g. due to a large strain burst), the crystal was completely unloaded then reloaded with the same procedure discussed above. The engineering stress was calculated by dividing the load by the initial mid-plane cross-sectional area, while the engineering strain was calculated by dividing the displacement by the microcrystal height. Before calculating strain, the displacement data was corrected to account for the machine compliance and the compliance due to the deformation of the base[72]. At least four tests were performed for microcrystals having $D$ = 2, 5, 10, 20 μm to guarantee repeatability of the observations.

**Molecular Dynamics Simulations**

All simulations were performed using the three-dimensional MD simulation code LAMMPS[73]. The embedded atom method (EAM) potential developed by Liu et al.[74] was used to model the Mg interatomic interaction. As summarized in ref.[75], this potential agree well with density function theory (DFT) and experimental measurements in terms of lattice constants, cohesive energy, elastic constants, and stacking fault energies. In particular, three rectangular simulation cells were modeled. The smallest simulation cell mimics a rectangular nano-crystal having edge-lengths $l_x$ = 30 nm, $l_y$ = 30 nm, and $l_z$ = 60 nm. The total number of atoms in this simulation cell is ~2.38 million atoms. The intermediate simulation cell mimics a cubic nano-crystal with edge length 100 nm, having ~48 million atoms. The largest simulation cell has edge lengths $l_x$ = 140 *nm*, $l_y$ = 140 *nm*, and $l_z$ = 70 *nm*, which contains ~59 million atoms. Free surface boundary conditions were employed along all three directions. The compressive loading was imposed along the $[11\bar{2}0]$ *a*-axis, which coincides with the *z*-direction of the simulation cell. The imposed strain rate was $5\times10^9$ s$^{-1}$ for the smallest and intermediate simulation cells and $2\times10^9$ s$^{-1}$, for the largest simulation cell.

At the beginning of all simulations, the atomic system was fully relaxed by the conjugate gradient algorithm with energy tolerance of $10^{-12}$ and force tolerance of $10^{-13}$ eV/A. The system was then heated to 300 K within 300 ps using the NVE ensemble, and the temperature was maintained constant during loading using the NVT ensemble. The time step was set to be 0.001 ps during loading for all simulations. In all atomic structure images reported here, thermal fluctuations were removed by performing 50 steps of conjugate gradient relaxation. All atom visualizations were obtained using OVITO[76], while the centro-symmetry and common neighbor analysis (CNA) parameters were used to color the atoms as indicated in each figure caption. The centro-symmetry parameter for HCP crystals are 0 for stacking faults or the FCC lattice; 9 for dislocation cores and TBs; 10 for HCP bulk lattice; and >12 for other.

**References**

1. Mordike BL, Ebert T. Magnesium - Properties - applications - potential. *Mat Sci Eng a-Struct* 2001, **302**(1): 37-45.

2. Caceres CH. Economical and environmental factors in light alloys automotive applications. *Metall Mater Trans A* 2007, **38A**(7): 1649-1662.

3. Kulekci MK. Magnesium and its alloys applications in automotive industry. *Int J Adv Manuf Tech* 2008, **39**(9-10): 851-865.

4. Bettles C, Barnett M. *Advances in wrought magnesium alloys : fundamentals of processing, properties and applications*. Woodhead Publishing: Oxford, UK, 2012.

5. Luo AA. Recent magnesium alloy development for elevated temperature applications. *Int Mater Rev* 2004, **49**(1): 13-30.

6. Yoo MH. Slip, Twinning, and Fracture in Hexagonal Close-Packed Metals. *Metall Trans A* 1981, **12**(3): 409-418.

7. Uchic MD, Dimiduk DM, Florando JN, Nix WD. Sample dimensions influence strength and crystal plasticity. *Science* 2004, **305**(5686): 986-989.

8. Dimiduk DM, Uchic MD, Parthasarathy TA. Size-affected single-slip behavior of pure nickel microcrystals. *Acta Mater* 2005, **53**(15): 4065-4077.

9. Greer JR, Oliver WC, Nix WD. Size dependence in mechanical properties of gold at the micron scale in the absence of strain gradients (vol 53, pg 1821, 2005). *Acta Mater* 2006, **54**(6): 1705-1705.

10. Kiener D, Motz C, Schoberl T, Jenko M, Dehm G. Determination of mechanical properties of copper at the micron scale. *Adv Eng Mater* 2006, **8**(11): 1119-1125.

11. Volkert CA, Lilleodden ET. Size effects in the deformation of sub-micron Au columns. *Philos Mag* 2006, **86**(33-35): 5567-5579.

12. Ng KS, Ngan AHW. Stochastic nature of plasticity of aluminum micro-pillars. *Acta Mater* 2008, **56**(8): 1712-1720.

13. Keller RM, Baker SP, Arzt E. Quantitative analysis of strengthening mechanisms in thin Cu films: Effects of film thickness, grain size, and passivation. *J Mater Res* 1998, **13**(5): 1307-1317.

14. Kraft O, Hommel M, Arzt E. X-ray diffraction as a tool to study the mechanical behaviour of thin films.

*Mat Sci Eng a-Struct* 2000, **288**(2)**:** 209-216.

15. Read DT, Cheng YW, Keller RR, McColskey JD. Tensile properties of free-standing aluminum thin films. *Scripta Materialia* 2001, **45**(5)**:** 583-589.

16. Haque MA, Saif MTA. Strain gradient effect in nanoscale thin films. *Acta Mater* 2003, **51**(11)**:** 3053-3061.

17. Espinosa HD, Prorok BC, Peng B. Plasticity size effects in free-standing submicron polycrystalline FCC films subjected to pure tension. *J Mech Phys Solids* 2004, **52**(3)**:** 667-689.

18. Yu DYW, Spaepen F. The yield strength of thin copper films on Kapton. *J Appl Phys* 2004, **95**(6)**:** 2991-2997.

19. Xiang Y, Vlassak JJ. Bauschinger and size effects in thin-film plasticity. *Acta Mater* 2006, **54**(20)**:** 5449-5460.

20. Gruber PA, Bohm J, Onuseit F, Wanner A, Spolenak R, Arzt E. Size effects on yield strength and strain hardening for ultra-thin Cu films with and without passivation: A study by synchrotron and bulge test techniques. *Acta Mater* 2008, **56**(10)**:** 2318-2335.

21. Sim GD, Park JH, Uchic MD, Shade PA, Lee SB, Vlassak JJ. An apparatus for performing microtensile tests at elevated temperatures inside a scanning electron microscope. *Acta Mater* 2013, **61**(19)**:** 7500-7510.

22. Sim GD, Vlassak JJ. High-temperature tensile behavior of freestanding Au thin films. *Scripta Materialia* 2014, **75:** 34-37.

23. Li JZ, Xu W, Wu XL, Ding H, Xia KN. Effects of grain size on compressive behaviour in ultrafine grained pure Mg processed by equal channel angular pressing at room temperature. *Mat Sci Eng a-Struct* 2011, **528**(18)**:** 5993-5998.

24. Barnett MR, Keshavarz Z, Beer AG, Atwell D. Influence of grain size on the compressive deformation of wrought Mg-3Al-1Zn. *Acta Mater* 2004, **52**(17)**:** 5093-5103.

25. Lapovok R, Thomson PF, Cottam R, Estrin Y. The effect of grain refinement by warm equal channel angular extrusion on room temperature twinning in magnesium alloy ZK60. *J Mater Sci* 2005, **40**(7)**:** 1699-1708.

26. Barnett MR. A rationale for the strong dependence of mechanical twinning on grain size. *Scripta Materialia* 2008, **59**(7)**:** 696-698.

27. Fan H, Aubry S, Arsenlis A, El-Awady JA. Grain size effects on dislocation and twinning mediated plasticity in magnesium. *Scripta Materialia* 2015.

28. Lilleodden E. Microcompression study of Mg (0001) single crystal. *Scripta Materialia* 2010, **62**(8)**:** 532-535.

29. Ye J, Mishra RK, Sachdev AK, Minor AM. In situ TEM compression testing of Mg and Mg-0.2 wt.% Ce single crystals. *Scripta Materialia* 2011, **64**(3)**:** 292-295.

30. Byer CM, Ramesh KT. Effects of the initial dislocation density on size effects in single-crystal magnesium. *Acta Mater* 2013, **61**(10)**:** 3808-3818.

31. Aitken ZH, Fan HD, El-Awady JA, Greer JR. The effect of size, orientation and alloying on the deformation of AZ31 nanopillars. *J Mech Phys Solids* 2015, **76:** 208-223.

32. Yu Q, Shan ZW, Li J, Huang XX, Xiao L, Sun J*, et al.* Strong crystal size effect on deformation twinning. *Nature* 2010, **463**(7279)**:** 335-338.

33. Kim GS, Yi S, Huang Y, Lilleodden E. Twining and Slip Activity in Magnesium <11-20> Single Crystal. *MRS Proceedings* 2009, **1224**.

34. Kim GS. Small volume investigation of slip and twinning in magnesium single crystals. PhD thesis, Université de Grenoble, Grenoble, 2011.

35. Yu Q, Mishra RK, Minor AM. The Effect of Size on the Deformation Twinning Behavior in Hexagonal Close-Packed Ti and Mg. *Jom-Us* 2012, **64**(10)**:** 1235-1240.

36. Yu Q, Qi L, Chen K, Mishra RK, Li J, Minor AM. The Nanostructured Origin of Deformation Twinning. *Nano Lett* 2012, **12**(2)**:** 887-892.

37. Ma Q, El Kadiri H, Oppedal AL, Baird JC, Li B, Horstemeyer MF*, et al.* Twinning effects in a rod-textured AM30 Magnesium alloy. *Int J Plasticity* 2012, **29:** 60-76.

38. El Kadiri H, Kapil J, Oppedal AL, Hector LG, Agnew SR, Cherkaoui M*, et al.* The effect of twin-twin interactions on the nucleation and propagation of {10(1)over-bar2} twinning in magnesium. *Acta Mater* 2013, **61**(10)**:** 3549-3563.

39. Yu Q, Wang J, Jiang Y, McCabe RJ, Tomé CN. Co-zone {1̄012} Twin Interaction in Magnesium Single Crystal. *Materials Research Letters* 2013, **2**(2)**:** 82-88.

40. Burke EC, Hibbard WR. Plastic Deformation of Magnesium Single Crystals. *T Am I Min Met Eng* 1952, **194**(3)**:** 295-303.

41. Conrad H, Robertson WD. Effect of Temperature on the Flow Stress and Strain-Hardening Coefficient of Magnesium Single Crystals. *T Am I Min Met Eng* 1957, **209:** 503-512.

42. Reedhill RE, Robertson WD. Deformation of Magnesium Single Crystals by Nonbasal Slip. *T Am I Min Met Eng* 1957, **209:** 496-502.

43. Yoshinaga H, Horiuchi R. Deformation mechanisms in magnesium single crystals compressed in the direction parallel to the hexagonal axis. *Trans JIM* 1963, **4**(1)**:** 1-8.

44. Kelley EW, Hosford WF. Plane-Strain Compression of Magnesium and Magnesium Alloy Crystals. *T Metall Soc Aime* 1968, **242**(1)**:** 5-&.

45. Obara T, Yoshinga H, Morozumi S. [112bar2](11bar23) Slip System in Magnesium. *Acta Metall Mater* 1973, **21**(7)**:** 845-853.

46. Xie KY, Alam Z, Caffee A, Hemker KJ. Pyramidal I slip in c-axis compressed Mg single crystals. *Scripta Materialia* 2015.

47. Li B, Ma E. Pyramidal slip in magnesium: Dislocations and stacking fault on the {1011} plane. *Philos Mag* 2009, **89**(14)**:** 1223-1235.

48. Yasi JA, Nogaret T, Trinkle DR, Qi Y, Hector LG, Curtin WA. Basal and prism dislocation cores in magnesium: comparison of first-principles and embedded-atom-potential methods predictions. *Model Simul Mater Sc* 2009, **17**(5).

49. Tang YZ, El-Awady JA. Formation and slip of pyramidal dislocations in hexagonal close-packed magnesium single crystals. *Acta Mater* 2014, **71:** 319-332.

50. Tang YZ, El-Awady JA. Highly anisotropic slip-behavior of pyramidal I < c plus a > dislocations in hexagonal close-packed magnesium. *Mat Sci Eng a-Struct* 2014, **618:** 424-432.

51. Fan H, El-Awady JA. Molecular Dynamics Simulations of Orientation Effects During Tension, Compression, and Bending Deformations of Magnesium Nanocrystals. *Journal of Applied Mechanics* 2015, **82**(10)**:** 101006-101006.

52. Fan H, El-Awady JA. Towards resolving the anonymity of pyramidal slip in magnesium. *Materials Science*

*and Engineering: A* 2015, **644:** 318-324.

53. Xie KY, Hemker KJ. *Unpublished work.*

54. Prasad KE, Rajesh K, Ramamurty U. Micropillar and macropillar compression responses of magnesium single crystals oriented for single slip or extension twinning. *Acta Mater* 2014, **65:** 316-325.

55. Prasad KE, Ramesh KT. In-situ observations and quantification of twin boundary mobility in polycrystalline magnesium. *Mat Sci Eng a-Struct* 2014, **617:** 121-126.

56. Fan H, Aubry S, Arsenlis A, El-Awady JA. The role of twinning deformation on the hardening response of polycrystalline magnesium from discrete dislocation dynamics simulations. *Acta Mater* 2015, **92:** 126-139.

57. Fan H, Aubry S, Arsenlis A, El-Awady JA. Discrete dislocation dynamics simulations of twin size-effects in magnesium.   MRS Proceedings; 2015; 2015.

58. Sun Q, Zhang XY, Ren Y, Tu J, Liu Q. Interfacial structure of {1 0 (1)over-bar 2} twin tip in deformed magnesium alloy. *Scripta Materialia* 2014, **90-91:** 41-44.

59. Zhu T, Li J, Samanta A, Leach A, Gall K. Temperature and strain-rate dependence of surface dislocation nucleation. *Phys Rev Lett* 2008, **100**(2).

60. Beyerlein IJ, Tome CN. A probabilistic twin nucleation model for HCP polycrystalline metals. *P Roy Soc a-Math Phy* 2010, **466**(2121)**:** 2517-2544.

61. Tome CN, Beyerlein IJ, Wang J, McCabe RJ. A multi-scale statistical study of twinning in magnesium. *Jom-Us* 2011, **63**(3)**:** 19-23.

62. Orowan E. *Dislocations and mechanical properties*. The American Institute of Mining and Metallurgical Engineers: New York, NY, 1954.

63. Thompson N, Millard DJ. Twin Formation in Cadmium. *Philos Mag* 1952, **43**(339)**:** 422-440.

64. Mendelson S. Zonal Dislocations and Twin Lamellae in Hcp Metals. *Mater Sci Eng* 1969, **4**(4)**:** 231-&.

65. Mendelson S. Dislocation Dissociations in Hcp Metals. *J Appl Phys* 1970, **41**(5)**:** 1893-&.

66. Vaidya S, Mahajan S. Accommodation and Formation of (1121) Twins in Co Single-Crystals. *Acta Metall Mater* 1980, **28**(8)**:** 1123-1131.

67. Wang J, Hirth JP, Tome CN. ((1)over-bar0 1 2) Twinning nucleation mechanisms in hexagonal-close-packed crystals. *Acta Mater* 2009, **57**(18)**:** 5521-5530.

68. Wang J, Hoagland RG, Hirth JP, Capolungo L, Beyerlein IJ, Tome CN. Nucleation of a ((1)over-bar 0 1 2) twin in hexagonal close-packed crystals. *Scripta Materialia* 2009, **61**(9)**:** 903-906.

69. Sasaki M, Marukawa K. Etch Pits at Dislocations in Magnesium Single Crystals. *Transactions of the Japan Institute of Metals* 1977, **18**(7)**:** 540-544.

70. Sulkowski B, Mikulowski B. Work Hardening of Magnesium Single Crystals Deformed to Stage B at Room Temperature. *Acta Phys Pol A* 2012, **122**(3)**:** 528-531.

71. Zhang H, Schuster BE, Wei Q, Ramesh KT. The design of accurate micro-compression experiments. *Scripta Materialia* 2006, **54**(2)**:** 181-186.

72. Sneddon IN. The relation between load and penetration in the axisymmetric boussinesq problem for a punch of arbitrary profile. *International Journal of Engineering Science* 1965, **3**(1)**:** 47-57.

73. Plimpton S. Fast Parallel Algorithms for Short-Range Molecular-Dynamics. *J Comput Phys* 1995, **117**(1)**:** 1-19.

74. Liu X-Y, Adams JB, Ercolessi F, Moriarty JA. EAM potential for magnesium from quantum mechanical forces. *Model Simul Mater Sci Eng* 1996, **4**(3)**:** 293.

75. Tang Y, El-Awady JA. Formation and slip of pyramidal dislocations in hexagonal close-packed magnesium single crystals. *Acta Mater* 2014, **71**(0)**:** 319-332.

76. Stukowski A. Visualization and analysis of atomistic simulation data with OVITO–the Open Visualization Tool. *Model Simul Mater Sci Eng* 2010, **18**(1)**:** 015012.

**Acknowledgments**

This research was sponsored by the Army Research Laboratory (ARL) grant #W911NF-12-2-0022, and in part by the National Science Foundation CAREER Award #CMMI-1454072. The views and conclusions contained here are those of the authors and should not be interpreted as representing the official policies, either expressed or implied, of ARL or the U.S. Government. The U.S. Government is authorized to reproduce and distribute reprints for Government purposes notwithstanding any copyright notation herein. G.-D.S. and J.A.E. also acknowledge fruitful discussions with, and the bulk scale stress-strain measurements from K.Y. Xie and K.J. Hemker at Johns Hopkins University.

**Author Contributions**

G.-D.S. and J.A.E. designed the experiments. G.-D.S., G.K, and S.L. fabricated microcrystals. G.-D.S. conducted mechanical experiments and material characterization. M.H.H., H.F., and J.A.E. conducted MD simulations. G.-D.S. and J.A.E. wrote the manuscript with input from all authors. All authors contributed to discussion of the results, provided input on the manuscript, and approved the final version.

**Competing Financial Interests**

The authors declare no competing financial interests.

Table I. Critical resolved shear stress (experiments), Peierls stress (MD simulations), and Schmid factor of the various deformation modes in Mg for compression loading along the $[11\bar{2}0]$ direction for the parent matrix, and $[2\bar{1}\bar{1}3]$ for the twinned region

| Slip system | Slip direction | CRSS (MPa) | Peierls stress (MPa) | | $M\ [11\bar{2}0]$ | $M\ [2\bar{1}\bar{1}3]$ |
| --- | --- | --- | --- | --- | --- | --- |
| | | | Edge | Screw | | |
| Basal | $\langle 11\bar{2}0 \rangle$ | $0.5^{40,\ 41,\ 43}$ | $0.3^{48}$ | $3.6^{48}$ | 0 | 0.44 |
| Prismatic | $\langle 11\bar{2}0 \rangle$ | $39^{42}$ | $13^{48}$ | $44^{48}$ | 0.43 | 0.12 |
| Pyramidal I | $\langle 11\bar{2}3 \rangle$ | $40^{45*}$ | $226^{50}$ | $15^{50}$ | 0.38 | 0.21 |
| Pyramidal II | $\langle 11\bar{2}3 \rangle$ | - | $208^{52}$ | $157^{52}$ | 0.44 | 0.4 |
| $\{10\bar{1}2\}$ Tensile twin | - | $12^{44}$ | - | - | 0.37 | 0.36 |

*The CRSS reported in ref.[45] is for second-order pyramidal slip, however, recent MD simulations[47, 49, 51, 52], and experimental slip trace analysis[46], indicate that this value coincides with Pyramidal I slip instead.

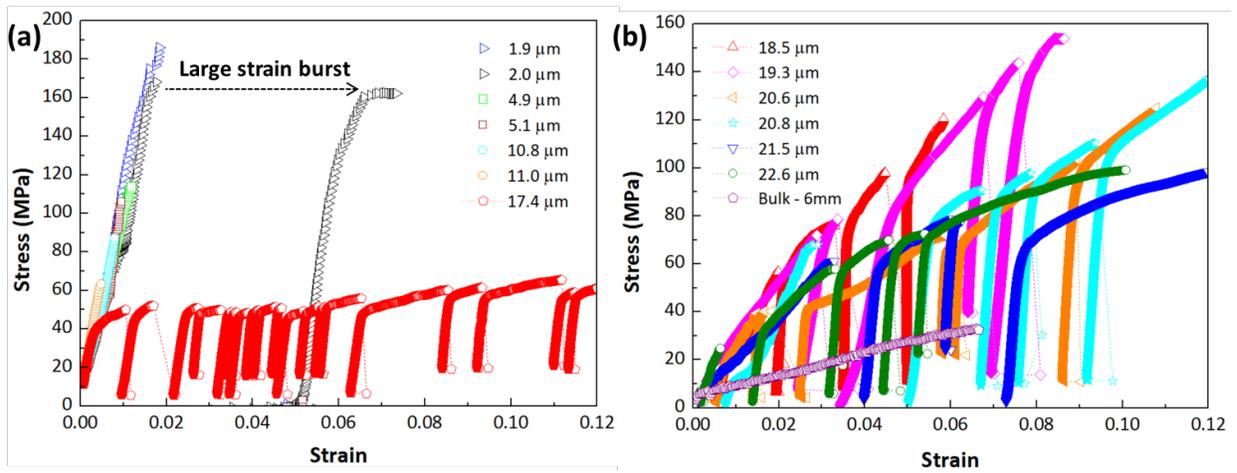

Fig.1. Representative stress-strain curves of a-axis oriented Mg microcrystals having sizes: (a) $D \leq 18$ μm; and (b) $D > 18$ μm. The stress-strain response of a bulk Mg single crystal compressed along the $[11\bar{2}0]$ direction is shown in (b) for comparison.

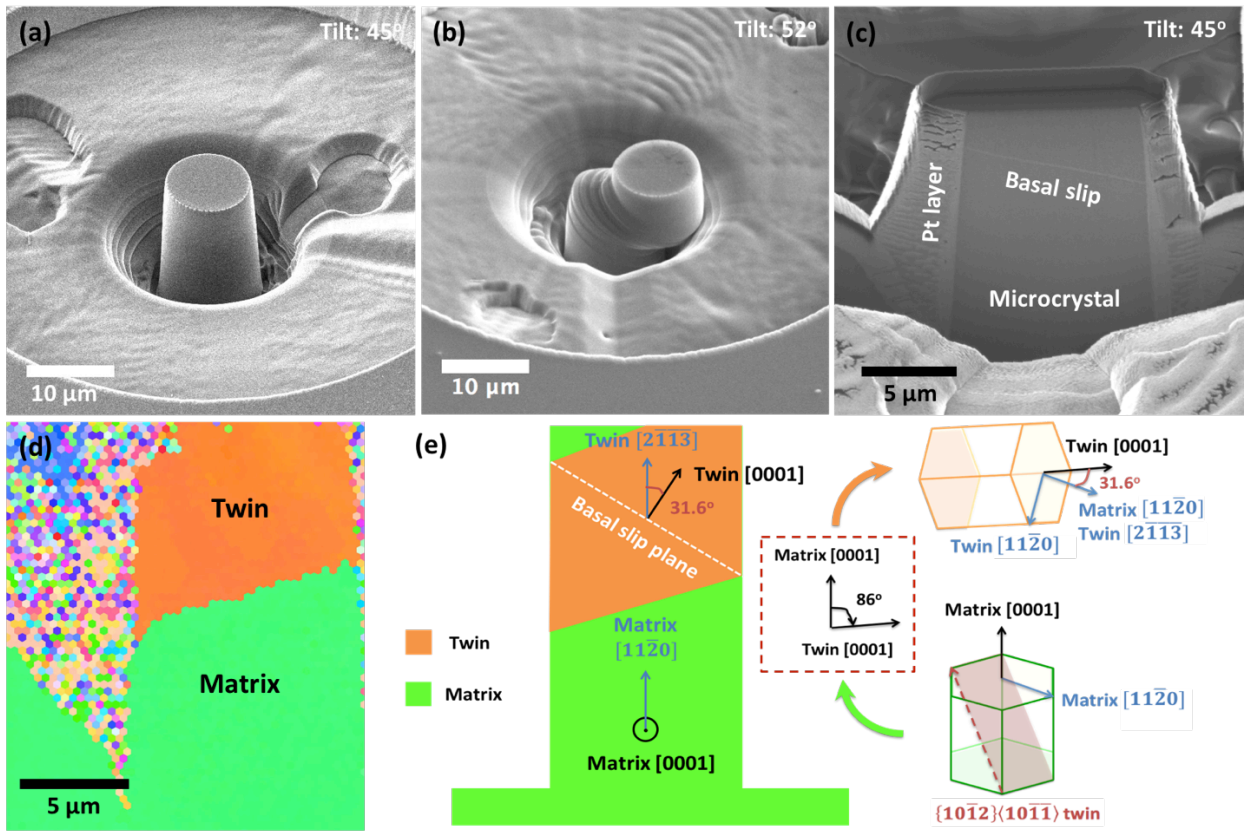

Fig.2. (a) SEM micrograph of a FIB milled Mg microcrystal; (b) SEM image of a *D* = 10 μm microcrystal showing massive basal slip bands after 15% strain. (c) Ion beam channeling contrast imaging of a thin foil extracted using FIB milling from the center of a *D* = 10 μm microcrystal after 5% strain. A basal slip band is clearly observed in the twinned region. (d) EBSD image of the cross-section in (c) showing a single twin propagated along the column height. (e) Schematic image of the pillar cross-section used to calculate the critical twin thickness required to activate a basal slip band, and crystal rotation showing how the loading axis changes after twinning.

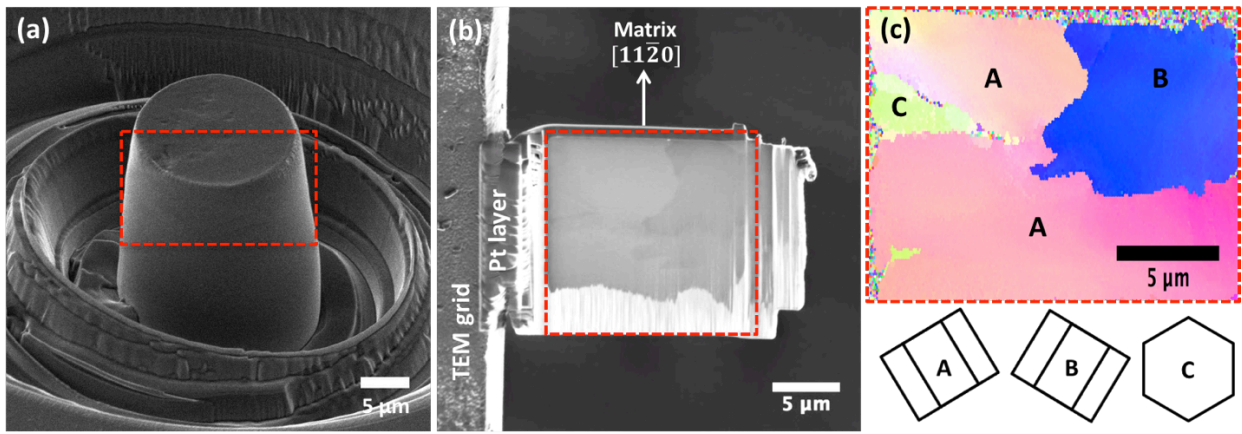

Fig.3. (a) SEM micrograph of a deformed 20 μm diameter microcrystal after 8% strain, showing barrel-like deformed geometry. The crystal initially had a 2:1 aspect ratio. (b) Electron channeling contrast imaging of a thin foil extracted from the top half at the center of the microcrystal shown in (a). (c) EBSD image of the thin foil shown in (b) revealing two twin-variants intersecting each other.

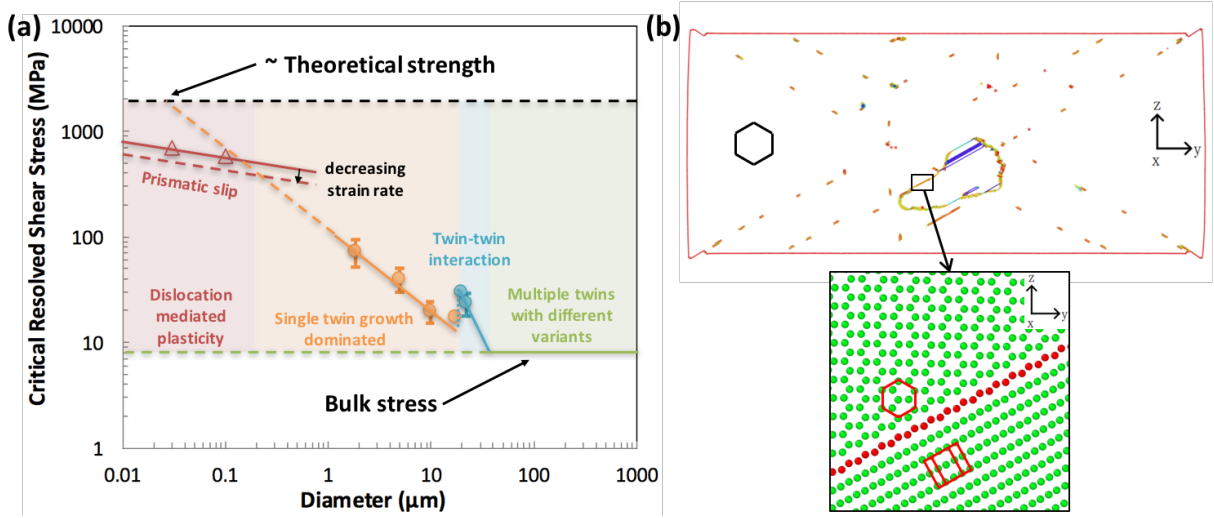

Fig. 4. (a) Critical resolved shear stress at 4% strain from experiments (circles), and at yield from MD simulations (deltas), versus microcrystal diameter. The deformation mechanism map is also shown by the colored regions. (b) Cross-section through the MD simulation cell for the 140 × 140 × 70 nm$^3$ simulation cell at 8.15% strain. All HCP atoms are removed to facilitate visualization of the defects. The insert shows detailed atomic structure of the boxed region. All atoms are colored according to their CNA parameter, where HCP atoms are green, FCC atoms are blue, and other atoms are red.

# Supporting Information

Anomalous Hardening in Magnesium Driven by a Size-Dependent Transition in Deformation Modes

Gi-Dong Sim[1], Gyuseok Kim[2], Steven Lavenstein[1], Mohamed H. Hamza[1], Haidong Fan[1,3], Jaafar A. El-Awady[1, *]

[1]Department of Mechanical Engineering, Johns Hopkins University, Baltimore, MD 21218, USA

[2]Singh Center for Nanotechnology, University of Pennsylvania, Philadelphia, PA, 19104, USA

[3]Department of Mechanics, Sichuan University, Chengdu, Sichuan 610065, China

* Email: jelawady@jhu.edu

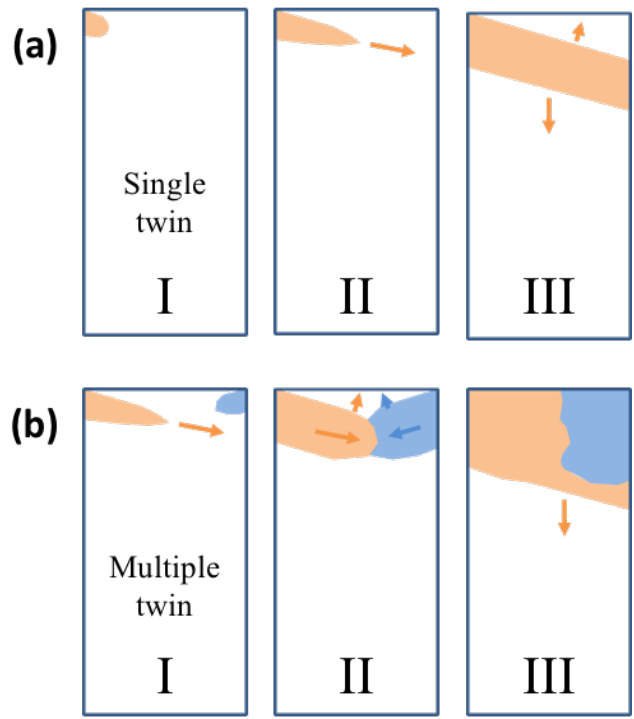

Fig. S1. Schematic image showing three steps during twin propagation for: (a) single twin mode; and (b) multiple twin mode.

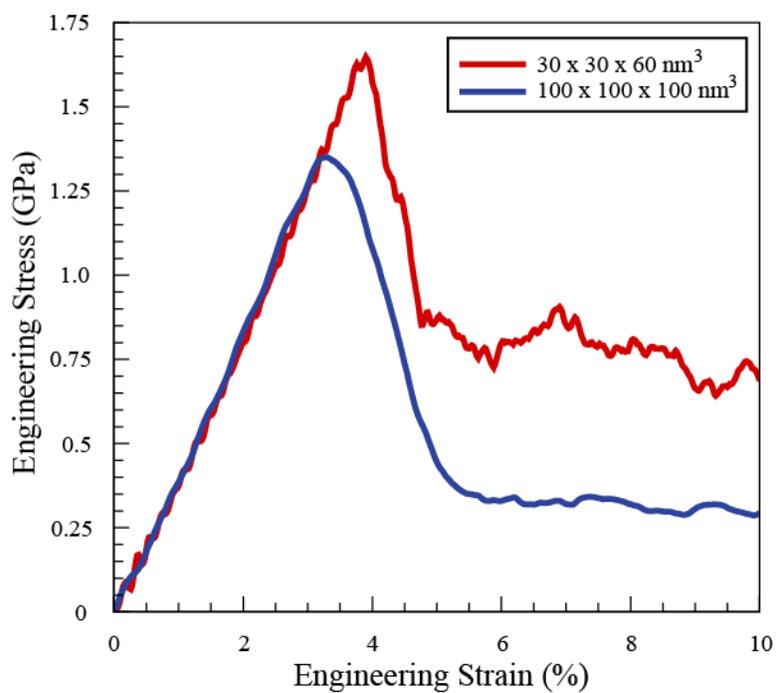

Fig. S2. Engineering stress-strain curves as predicted from MD simulations of two rectangular simulation cells oriented for compression along the $[11\bar{2}0]$ direction.

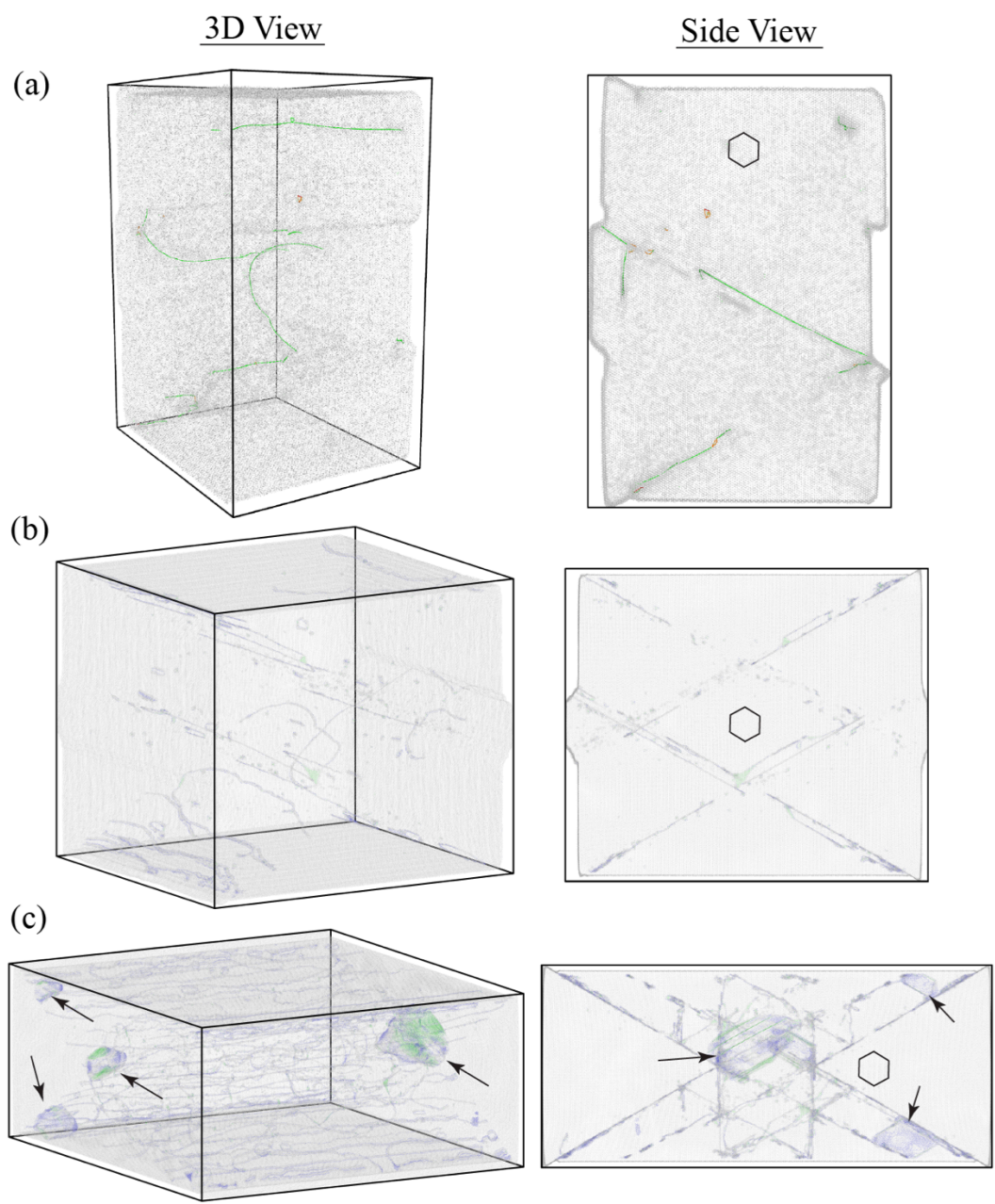

Fig. S3. Three-dimensional and side views of the atomic microstructures in three rectangular simulation cells: (a) 30 × 30 × 60 nm$^3$ cell at 16.04% strain; (b) 100 × 100 × 100 nm$^3$ cell at 7.68% strain; and (c) 140 × 140 × 70 nm$^3$ cell at 8% strain. All HCP atoms are removed to facilitate visualization. The arrows in (c) indicate twinned regions.

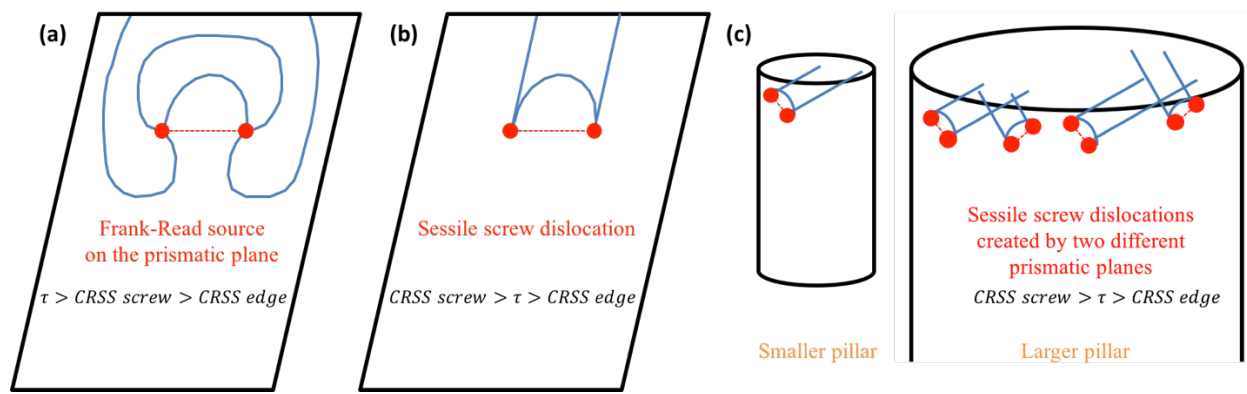

Fig. S4. Schematic image showing a (a) Frank-Read source being fully activated on the prismatic plane when the shear stress is higher than CRSS of both screw and edge dislocations; (b) Sessile screw dislocation forming when the shear stress is lower than the CRSS of screw dislocations; (c) Size effect on screw dislocation formation. Smaller crystals are expected to have limited number of source available to create sessile dislocations.